\documentclass[pra,aps,twocolumn,superscriptaddress,floatfix,noshowpacs,nofootinbib,10pt]{revtex4-2}

\usepackage[utf8]{inputenc}     	
\usepackage{amsmath, amssymb, mathtools, mathrsfs}
\usepackage[colorlinks=true]{hyperref}
\hypersetup{colorlinks,linkcolor={red},citecolor={blue},urlcolor={blue}}  
\usepackage{color}
\usepackage{graphicx}
\usepackage[normalem]{ulem}
\usepackage{mathtools}
\usepackage{stmaryrd}
\usepackage{float}
\usepackage{textalpha}
\usepackage{amsmath}

\begin{document}
\title{Phonon scattering from spatial relaxation of one-dimensional Bose gases}

\author{Bilal Alilou}
\email{bilal.alilou@lkb.upmc.fr}
\affiliation{Laboratoire Kastler Brossel, Sorbonne Universit\'{e}, CNRS, ENS-PSL Research University, 
Coll\`{e}ge de France; 4 Place Jussieu, 75005 Paris, France}

\author{Cl\'ement Duval}
\email{clement.duval2@cea.fr}
\affiliation{CEA DAM-DIF, F-91297 Arpajon, France}
\affiliation{Universit\'e Paris-Saclay, CEA, LMCE, F-91680 Bruy\`eres-le-Ch\^atel, France}

\author{Frederick Del Pozo}
\email{frederick.delpozo@lkb.upmc.fr}
\affiliation{Laboratoire Kastler Brossel, Sorbonne Universit\'{e}, CNRS, ENS-PSL Research University, 
Coll\`{e}ge de France; 4 Place Jussieu, 75005 Paris, France}

\author{Nicolas Cherroret}
\email{nicolas.cherroret@lkb.upmc.fr}
\affiliation{Laboratoire Kastler Brossel, Sorbonne Universit\'{e}, CNRS, ENS-PSL Research University, 
Coll\`{e}ge de France; 4 Place Jussieu, 75005 Paris, France}

\begin{abstract}
We theoretically investigate the nonequilibrium relaxation of a spatial density modulation in a one-dimensional, weakly interacting Bose gas, and its connection to the equilibrium  scattering rate $\smash{\gamma_k\propto k^{3/2}}$ of the system's phononic excitations.
We show that the relaxation is generally governed by a nonequilibrium scattering rate $\gamma_{k,t}$ coupled to quantum fluctuations, which approaches its equilibrium value $\gamma_k$ only at long times. Numerical simulations of quantum kinetic equations reveal an algebraic convergence, $\smash{\gamma_{k,t} - \gamma_k \sim t^{-2/3}}$, confirmed by analytical predictions. More broadly, our results establish a theoretical framework for experimentally probing phonon dynamics through the temporal evolution of local perturbations in quantum gases.
\end{abstract}
\maketitle

\section{Introduction} 

As the fundamental collective excitations of quantum gases, phonons play a key role in understanding the low-energy behavior of Bose-Einstein condensates \cite{Stringari_pitaevskii2003},  or quasi-condensates in low dimensions \cite{Popov1972, Mora2003}. Their linear sound-wave dispersion $\epsilon_k = c|k|$, with $c$ the speed of sound, underlies not only the phenomenon of superfluidity, but also more broadly the hydrodynamic properties of ultracold Bose gases. 
Theoretically, phonons naturally emerge in the Bogoliubov description of dilute gases~\cite{Stringari_pitaevskii2003}, as well as within the more general Luttinger liquid framework \cite{Protopopov2014, Cazalilla2004, Cazalilla2011}. In both formalisms, phonons are treated as quasiparticles of infinite lifetime — an approximation that breaks down once interactions between  quasiparticles are taken into account.

In Bose gases, phonons interact via the so-called Landau and Beliaev scattering processes \cite{Beliaev1958, Pitaevskii1997, Giorgini1998, Bighin2015, Micheli2022}, which endow them with a finite lifetime given by an inverse phonon scattering rate $\gamma_k$. In three-dimensional dilute gases at finite temperature, the Fermi Golden rule predicts that $\gamma_k\propto k$ \cite{Pitaevskii1997, Chung2009}. A similar result holds in two dimensions, albeit with a prefactor renormalized by nonperturbative corrections~\cite{Castin2023}. 
The behavior of phonon scattering is, in contrast, markedly different in one-dimensional (1D) Bose gases where  $\gamma_k\propto k^{3/2}$ exhibits a nonanalytic infrared scaling. This characteristic $3/2$ exponent, first derived by Andreev \cite{Andreev1980}, has attracted considerable attention as it is believed to be connected with the possible existence of a Kardar–Parisi–Zhang (KPZ) dynamics of density fluctuations in the low-energy regime \cite{Beijeren2012, Kulkarni2013, Kulkarni2015}. 

Experimentally, direct measurements of phonon scattering processes remain scarce. Their signatures have nevertheless been observed through Bragg spectroscopy~\cite{Katz2002} or in out-of-equilibrium configurations involving, for instance, the spatial propagation of sound waves \cite{Ville2018} or the evolution of structure factors after interaction quenches in Bose gases \cite{Hung2013}. As a matter of fact, in quantum quench experiments, phonon scattering constitutes the microscopic mechanism responsible for the eventual re-thermalization of a Bose gas after the initial perturbation \cite{Tavora2013, Regemortel2018, Duval2023, Duval2025}.
In weakly interacting gases, this relaxation proceeds slowly, giving rise to a ``prethermal'' regime where quasiparticles hardly interact, before full thermalization sets in \cite{Gring2012, Langen2013, Larre2018, Abuzarli2022}. 

To experimentally determine the phonon lifetime, a natural strategy is to imprint a local perturbation to an initially homogeneous Bose gas and monitor its subsequent relaxation, for instance through the time evolution of the spatial density. At sufficiently late times, once the gas has nearly returned to equilibrium, this relaxation is expected to become purely exponential with a decay rate set by $\gamma_k$. The key questions are then how long the dynamics must be tracked for such a measurement to accurately capture $\gamma_k$, and to what extent the relaxation is genuinely exponential. These are the central issues we address theoretically in this work, focusing on the case of one-dimensional Bose gases.

In detail, we theoretically investigate how a weak and periodic density modulation of wavelength $2\pi/k_0$, initially imprinted on a 1D Bose gas, relaxes  until full thermalization is  reached. 
This modulation is described in terms of a nonzero quasiparticle mean field, which evolves in time alongside the system's quantum fluctuations encoded in the phonon momentum distribution.
Within this framework, we derive an explicit expression for the spatial density at arbitrary times after the quench, and show that its relaxation is controlled by a \emph{time-dependent} scattering rate $\gamma_{k_0,t}$ coupled to quantum fluctuations, which only approaches its equilibrium value $\smash{\gamma_{k_0}}$ at late times.
Extensive numerical simulations of quantum kinetic equations reveal that this convergence is rather slow, following an algebraic scaling $\gamma_{k_0,t} - \gamma_{k_0} \sim t^{-2/3}$, which is supported by analytical arguments. Finally, we demonstrate that the relaxation time associated with $\gamma_{k_0,t}$ is itself of the order of the equilibrium  phonon lifetime $\gamma_{k_0}^{-1}$.
\newpage
The paper is organized as follows. In Sec.~\ref{Sec:hydro}, we introduce the quantum hydrodynamic framework for Bose gases, which forms the basis of our nonequilibrium description. In Sec.~\ref{Sec:quench_protocol}, we describe a realistic nonequilibrium protocol --- a brief periodic-potential quench applied to a 1D Bose gas --- and derive the resulting post-quench state. This serves as the initial condition for the subsequent density dynamics.
The theoretical description of this dynamics is presented in Sec.~\ref{Sec:post_quench_dyn}, while the underlying Keldysh field-theoretic derivation is deferred to Sec.~\ref{Sec:FieldTheory} for clarity. Our main results, in particular the time evolution of the phonon scattering rate, are presented in Sec.~\ref{Sec:results}. Finally, Sec.~\ref{Sec:Conclusion} summarizes our findings and outlines perspectives for future work.

\section{Hydrodynamics of 1D Bose gases}
\label{Sec:hydro}

Consider a homogeneous, 1D Bose gas of $N$ identical bosons of mass $m$. We denote by $\rho_0 = N/L$ its mean density, where $L$ is the system size. From now on, we focus on the thermodynamic limit $N\to\infty$, $L\to\infty$. The bosons interact with each other through a two-body repulsive contact potential of strength $g>0$. The second-quantized, many-body Hamiltonian reads
\begin{equation}
    \hat{H} = \int dx\left(-\frac{1}{2m}\hat{\Psi}^\dagger\partial^2_x\hat{\Psi}+\frac{g}{2}\hat{\Psi}^\dagger\hat{\Psi}^\dagger\hat{\Psi}\hat{\Psi}\right),
    \label{eq:microH}
\end{equation}
in units of $\hbar = 1$.
We focus on the relaxation dynamics of a weakly interacting Bose gas initially at equilibrium and at very low temperature, within the “quasi-condensate” regime \cite{Bouchoule2011}. To describe this system, we make use of a hydrodynamic framework in which we express the field operator $\hat\Psi$ as
\begin{equation}
    \hat{\Psi}(x) = e^{i\hat{\theta}(x)}\sqrt{\hat{\rho}(x)},
    \label{eq:DP}
\end{equation}
and write the density operator as $\hat{\rho}(x)=\rho_0+\delta\hat{\rho}(x)$. The density-fluctuation and phase operators satisfy the commutation relation $[\delta \hat{\rho}(x), \hat{\theta}(x')] = i\delta(x-x')$.
In what follows, $\delta\hat{\rho}$ and $\partial_x\hat{\theta}$ will encompass \emph{both} the intrinsic quantum and thermal fluctuations of the Bose gas, as well as the externally imposed modulation at initial time. All these fluctuations are assumed small, which implies that $|\delta\hat{\rho}(x)|,|\partial_x\hat\theta| \ll \rho_0$ and allows us to expand Eq.~\eqref{eq:microH} perturbatively. To third order, we obtain the hydrodynamic Hamiltonian
\begin{equation}\begin{aligned}
    \hat{H} &\simeq& \int dx~\Big[\frac{\rho_0}{2m}(\partial_x\hat{\theta})^2+\frac{g}{2}\left(\delta\hat{\rho}\right)^2+\frac{1}{8m\rho_0}(\partial_x\delta\hat{\rho})^2\\
    &&+\frac{1}{2m}(\partial_x\hat{\theta})\delta\hat{\rho}(\partial_x\hat{\theta})\Big].
    \label{eq:hydroH}
\end{aligned}\end{equation}
The next step is to move to the quasiparticle basis by diagonalizing the quadratic part of $ \hat{H}$. This is achieved through the Bogoliubov transformation:
\begin{equation}
    \delta\hat{\rho}_{k} = -\sqrt{s_k}(\hat{a}_{k}^\dagger+\hat{a}_{-k}^{\phantom{\dagger}})~\mbox{,}~\hat{\theta}_{k} = \frac{i}{2\sqrt{s_k}}(\hat{a}_{k}^\dagger-\hat{a}_{-k}^{\phantom{\dagger}}),
    \label{eq:Bogoliubov}
\end{equation}
where we introduced the Fourier transforms
\begin{equation}
    \hat{\theta}_{k}\equiv\rho_0\int dx~\hat{\theta}(x)e^{-ikx}~\mbox{,}~\delta\hat{\rho}_{k}\equiv\int dx~\delta\hat{\rho}(x)e^{-ikx},
    \label{eq:FourierTransform}
\end{equation}
and where we defined $s_k \equiv E_k/\epsilon_k$, with $E_k\equiv k^2/(2m)$  and $\epsilon_k \equiv \sqrt{E_k(E_k+2g\rho_0)}$ the free-particle and Bogoliubov dispersion relations, respectively. The  annihilation operators $\hat{a}_{k}$ obey the usual canonical commutation relations of bosons, and define the quasiparticles of the weakly interacting Bose gas. In the following, it will be sufficient to focus on their low-energy behavior, where $\epsilon_k\simeq c|k|$ takes the form of a phononic dispersion with speed of sound $\smash{c \equiv \sqrt{g\rho_0/m}}$. In this limit, the Hamiltonian \eqref{eq:hydroH} becomes
\begin{equation}
    \hat{H}\! =\! \int_{k}\epsilon_k\left(\hat{a}^\dagger_{k}\hat{a}_{k}^{\phantom{\dagger}}
    \!+\!\frac{1}{2}\right)
\!+\! \int_{k,p}\!\Lambda_{k,p}\left(\hat{a}_{p}^{\phantom{\dagger}}\hat{a}_{k}^{\phantom{\dagger}}\hat{a}^\dagger_{p+k}\!+\!\text{h.c.}\right),
    \label{eq:Hhydroa}
\end{equation}
where $\smash{\int_k\equiv \int_{-\infty}^\infty \mathrm{d}k/(2\pi\rho_0)}$ and the coefficient of the cubic terms is $\Lambda_{p,k} \equiv 3/(4m)\sqrt{g\rho_0|p||k||p+k|/(2c)}$. In the following, we investigate the nonequilibrium dynamics of the 1D Bose gas on the basis of the effective, low-energy model defined by Eq. \eqref{eq:Hhydroa}. 
\section{Quench protocol}
\label{Sec:quench_protocol}

\begin{figure}
\centering
\includegraphics[width=0.9\linewidth]{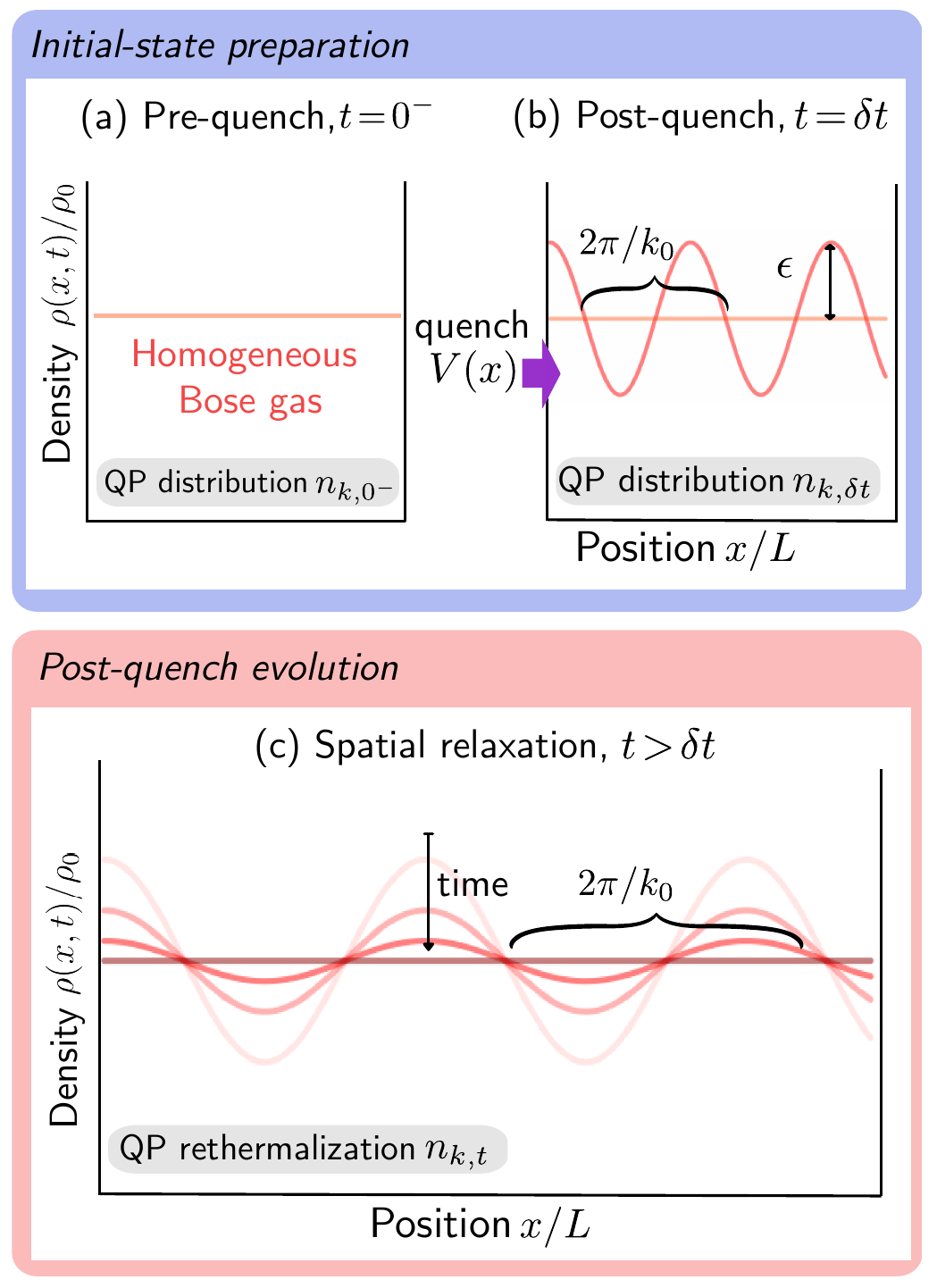}
\caption{\label{Fig_protocol}
Quench protocol and post-quench evolution of the spatial density. (a) Before the quench, the gas is homogeneneous, with a phonon distribution given by Eq. (\ref{eq:npprequench}). At $t=0$, a periodic potential is switched on, and switched off shortly after, at $t=\delta t$. (b) Right after the potential quench ($t=\delta t$), the density profile has acquired a periodic modulation, Eq.~(\ref{eq:density_a}), and the phonon distribution is given by Eq.~(\ref{eq:pq_mom}).
(c)~At later time, the density modulation exhibits both coherent oscillations and damping, cf. Eq.~(\ref{eq:rhofinal}). The figure illustrates the effect of damping at times integer multiples of $2 \pi/\epsilon_{k_0}$.
} 
\end{figure}
In this work, we aim to characterize the relaxation dynamics of a small perturbation initially imprinted on the spatial density profile of a 1D Bose gas, and to determine to what extent this relaxation provides access to the (equilibrium) phonon scattering rate. 
To this end, we consider the non-equilibrium protocol  illustrated in Fig. \ref{Fig_protocol}. Initially (at time $t=0^-$), the Bose gas is homogeneous and in thermal equilibrium at temperature $T_0$ [Fig. \ref{Fig_protocol}(a)]. Its mean density is
\begin{equation}\begin{aligned}
\rho(x,t=0^-)\equiv\langle\hat\Psi^\dagger(x,0^-)\hat\Psi(x,0^-)\rangle=\rho_0,
\end{aligned}\end{equation}
and its quasiparticle momentum distribution is given by
\begin{equation}\begin{aligned}
\label{eq:npprequench}
n_{k,t=0^-}\equiv\langle \hat a^\dagger_{k,0^-}\hat a_{k,0^-}^{\phantom{\dagger}}\rangle=\frac{1}{\exp(\epsilon_k/T_0)-1},
\end{aligned}\end{equation}
where we employ the Heisenberg representation for time-dependent operators.
As a realistic model for the spatial perturbation, we assume that a brief periodic potential $V(x)=V_0\cos(k_0x)$, of wavelength $2\pi/k_{0}$, is suddenly switched on and off at $t=0$ and $t = \delta t$, respectively. This quench imprints a corresponding density modulation on the gas [Fig. \ref{Fig_protocol}(b)], which then relaxes once the potential is removed, for times $t>\delta t$ [Fig. \ref{Fig_protocol}(c)].

Anticipating the next section, we note that for a weakly interacting gas, the dynamics of the spatial density is fully determined by the coupled evolutions of the first two cumulants of the phonon operator, i.e., by $\langle\hat a_{k, t}\rangle$ and $\smash{n_{k, t}\equiv \langle\hat a^\dagger_{k,t}\hat a_{k,t}^{\phantom{\dagger}}\rangle_c\equiv \langle\hat a^\dagger_{k,t}\hat a_{k,t}^{\phantom{\dagger}}\rangle-\langle\hat a^\dagger_{k,t}\rangle\langle \hat a_{k,t}^{\phantom{\dagger}}\rangle}$ (the connected correlator). 
To describe the post-quench dynamics, it is therefore sufficient to determine the value of these quantities at $t=\delta t$. 
The post-quench quasiparticle operator can be computed using the Heisenberg equation of motion
\begin{equation}
\label{eq:Heom}
    \partial_t\hat{a}_{k,t} = i[\hat{H},\hat{a}_{k,t}],
\end{equation}
where the full quench Hamiltonian is now given by the sum $\hat{H} = \hat{H}_0+\hat{H}_{\text{int}}+\hat{H}_{\text{pot}}$. Here, $\hat{H}_0$ and $\hat{H}_{\text{int}}$ respectively denote the quadratic and cubic parts in the right-hand side of Eq. \eqref{eq:Hhydroa}, while $\smash{\hat{H}_{\text{pot}}}$ represents the additional periodic potential in the quasiparticle basis:
\begin{align}
    \hat{H}_{\text{pot}} &\equiv \int dx \,V(x) \delta\hat\rho(x)\nonumber\\
    &= -\frac{V_0}{2}\sqrt{s_{k_0}}(\hat{a}^\dagger_{k_0}+\hat{a}_{k_0}^{\phantom{\dagger}}+\hat{a}^\dagger_{-k_0}+\hat{a}_{-k_0}^{\phantom{\dagger}}).
\end{align}
Suppose that, as a first approximation, we neglect the cubic part $H_\text{int}$ of the Hamiltonian during the quench.  Then Eq. (\ref{eq:Heom}) can be solved exactly, yielding 
\begin{align}
    \hat{a}_{k,\delta t} =\langle\hat{a}_{k,\delta t}\rangle  +\hat{a}_{k,0}e^{-i\epsilon_{k_0}\delta t}
    \label{eq:apostquench},
\end{align}
where
\begin{align}
    \langle\hat{a}_{k,\delta t}\rangle =& 
    \frac{\pi V_0}{2g\sqrt{s_{k_0}}}
    [\delta(k\!-\!k_0)\!+\!\delta(k\!+\!k_0)]
    (1\!-\!e^{-i\epsilon_{k_0}\delta t}).
     \label{eq:meanapostquench}
\end{align}
The first term on the right-hand side of Eq.~(\ref{eq:apostquench}) is a finite number, showing that, following the application of the periodic potential, the Bogoliubov operator acquires a \emph{non-zero} expectation value.  This unusual feature arises from our treatment of the spatial modulations, which are encoded in density and phase fluctuations rather than in an inhomogeneous background. 
Consequently, the mean density fluctuations $\langle\delta\hat\rho\rangle$ of the gas, related to $\langle \hat{a}_k \rangle$ via Eq.~\eqref{eq:Bogoliubov}, are also finite. The expression (\ref{eq:meanapostquench}) for the quasiparticle mean field is valid as long as the quench is weak, i.e., $V_0\ll g\rho_0$. 
On the other hand, the second term on the right-hand side of Eq. (\ref{eq:apostquench}) describes the evolution of quasiparticle \emph{fluctuations} during the quench.  From Eq. (\ref{eq:apostquench}), it follows that $n_{k, \delta t}\equiv \langle\hat a^\dagger_{k,\delta t}\hat a_{k,\delta t}^{\phantom{\dagger}}\rangle_c= n_{k,0^-}$ at this level of approximation.
In other words, neglecting $\smash{\hat{H}_\text{int}}$ during the quench leads to no change in the quasiparticle momentum distribution: at quadratic order, there is no coupling between the quasiparticle mean field and the quasiparticle fluctuations.

Since our goal is to describe a more generic situation in which the quench affects both the mean field and its fluctuations, we now consider the effect of $\smash{\hat{H}_\text{int}}$ during the quench. At leading order in $V_0/g\rho_0\ll1$ and for a short quench $\epsilon_{k_0}\delta t\ll1$, Eq. (\ref{eq:meanapostquench}) for the mean field remains valid but the post-quench quasiparticle distribution becomes (see Appendix \ref{Sec:appendixA} for details)
\begin{equation}
    n_{k,\delta t} = n_{k,0^-}
    \!+\! \ \frac{\varepsilon^2|k|}{k_0^2}\left[|k-k_0|n_{k-k_0,0^-}\!+\!|k+k_0|n_{k+k_0,0^-}\right]
    \label{eq:pq_mom}
\end{equation}
where $\varepsilon \equiv 3V_0(\epsilon_{k_0}\delta t)^2/(16g\rho_0)$. 
Thus, after the quench, the phonon distribution \eqref{eq:pq_mom} develops a double-peak structure at $k=\pm k_0$, superimposed on the pre-quench thermal background $n_{k,0^-}$. These peaks are not sharply localized at $k=\pm k_0$, but are instead broadened due to phonon interactions.

Together, Eqs. \eqref{eq:meanapostquench} and \eqref{eq:pq_mom} characterize a nonequilibrium post-quench state, which subsequently evolves for times $t > \delta t$. The next section is devoted to the theoretical modeling of this evolution.

\section{Post-quench evolution of density}
\label{Sec:post_quench_dyn}

\subsection{Bogoliubov approximation}

From now on we redefine for clarity the origin of time as the post-quench time: $t=\delta t\to t=0$. To describe the post-quench evolution of the density, the simplest approach is to  completely neglect phonon–phonon interactions [i.e., omit the second term on the right-hand side of Eq.~\eqref{eq:Hhydroa}]. Within this ``Bogoliubov approximation'', the Hamiltonian is purely quadratic and the phonons constitute the exact eigenmodes of the system. Consequently, the quasiparticle operator evolves purely harmonically, $\hat a_{k,t}=\hat a_{k,0}\exp(-i\epsilon_k t)$. This implies that the out-of-equilibrium momentum distribution \eqref{eq:pq_mom} induced by the quench does not evolve in time, and that the mean density is given by:
\begin{align}
\label{eq:density_a}
    \rho(x,t)\!&\equiv\! \rho_0
    \left[1-\int_k\sqrt{s_k}(\langle\hat{a}_{k, t}^\dagger\rangle+
    \langle\hat{a}_{-k,t}^{\phantom{\dagger}}\rangle)e^{ikx}\right]\\    &=\rho_0\left[1\!-\!\frac{V_0\epsilon_{k_0}\delta t}{g\rho_0}\cos(k_0x)
    \sin[\epsilon_{k_0}(t+\delta t/2)]
    \right]\nonumber
\end{align}
where we used Eq.~\eqref{eq:Bogoliubov} in the first equality, and Eq.~\eqref{eq:meanapostquench} in the second, assuming $\epsilon_{k_0}\delta t\ll1$.
The spatial modulation imprinted by the external potential at $t=0$ thus causes the density to evolve as a standing wave at later times. This oscillation can be viewed as the superposition of two counter-propagating sound waves with momenta $\pm k_0$  generated by the quench. Within the Bogoliubov approximation, phonons have an infinite lifetime, and the propagating sound waves remain undamped.
\subsection{Density relaxation}
\label{Sec:density_relax}
We now turn to the full scenario where phonon–phonon interactions are accounted for after the quench. These interactions cause the phonons to acquire a complex energy shift, represented by a (retarded) self-energy $\Sigma^R_{x,k,t}$. 
Since after the quench the gas is both out of equilibrium and spatially inhomogeneous, the self-energy here depends on position $x$, momentum $k$ and time $t$. 
Using a Keldysh field  theory--- presented separately in Sec. \ref{Sec:FieldTheory} for clarity--- we find that the expectation value of the Bogoliubov operator now takes the form
\begin{equation}
    \langle\hat{a}_{k,t}\rangle = \langle\hat{a}_{k,0}\rangle 
    \exp\left[-i\epsilon_kt-i\int_x\int_{0}^{t}dt'\Sigma^R_{x,k,t'}\right]
    \label{eq:Akt}
\end{equation}
with $\int_x\equiv\int dx/L$. Because the self-energy has a finite imaginary part, this relation ultimately leads to the relaxation of the density, $\rho(x,t\to\infty)=\rho_0$, which is related to $ \langle\hat{a}_{k,t}\rangle$ through the first line of Eq.~\eqref{eq:density_a}. 
The description of this relaxation thus boils down to the knowledge of $\Sigma^R_{x,k,t}$. 
For weakly interacting gases at \emph{equilibrium}, the self-energy was investigated in previous works. In three-dimensional Bose gases, it can be computed within the Born approximation, or equivalently via the Fermi golden rule \cite{Pitaevskii1997}. 
In two dimensions, the Born approximation still provides qualitatively correct results \cite{Chung2009, Duval2023, Duval2025}, although a quantitative description requires the inclusion of higher-order corrections \cite{Castin2023}.
In contrast, for 1D Bose gases at equilibrium, the Born approximation predicts an exact vanishing of the self-energy and is therefore insufficient \cite{Lange2012}.
This failure of leading-order perturbation theory originates from the non-analytic infrared behavior of the self-energy at low momenta in one dimension, and can be remedied by employing a self-consistent calculation \cite{Andreev1980}. 

In the present scenario, where the 1D Bose gas is both \emph{out of equilibrium} and \emph{inhomogeneous}, a time-dependent and local self-consistent treatment must be employed. This methodology, which was previously used in the context of homogeneous Luttiger liquids \cite{Buchhold2015, Buchhold2016}, is detailed in Sec. \ref{Sec:FieldTheory} for clarity. The resulting expression for $\Sigma^R_{x,k,t}$ satisfies the following self-consistent relation 
\begin{equation}\begin{aligned}
    \Sigma^R_{x,k,t} =& -2\int_0^\infty\frac{\mathrm{d}p}{\pi\rho_0}~\frac{\Lambda_{k,p}^2(n_{x,p,t}-n_{x,k+p,t})}{\Sigma^R_{x,k+p,t}-\Sigma^{R*}_{x,p,t}-\Sigma^{R*}_{x,k,t}}\\
    &-\int_0^{k}\frac{\mathrm{d}p}{\pi\rho_0}~\frac{\Lambda_{p,k-p}^2(n_{x,k-p,t}+n_{x,p,t}+1)}{\Sigma^R_{x,p,t}+\Sigma^R_{x,k-p,t}-\Sigma^{R*}_{x,k,t}},
    \label{eq:self-energyR}
\end{aligned}\end{equation}
for $k>0$, with the property $\Sigma^R_{x,-k,t}=\Sigma^R_{x,k,t}$. In this relation,
 $n_{x,k,t}$ is the phonon Wigner distribution, defined from the quasiparticle correlator as:
\begin{equation}
    n_{x,k,t} \equiv 
    N\int_{k'}
    e^{ik'x}
    \langle\hat{a}^\dagger_{k+\frac{k'}{2},t}\hat{a}_{k-\frac{k'}{2},t}^{\phantom{\dagger}}\rangle_c.
    \label{eq:wignerphonon}
\end{equation}
As the Wigner distribution is a real quantity, Eq.~\eqref{eq:self-energyR} implies that $\Sigma^R_{x,k,t}$ is purely imaginary : $\Sigma^{R*}_{x,k,t}=-\Sigma^R_{x,k,t}$ \cite{Buchhold2015, Buchhold2016}.
Evaluating the self-energy at a given time thus requires the knowledge of the Wigner distribution $n_{x,k,t}$ at the same time. This information is provided by a quantum kinetic equation describing phonon-phonon collisions. This equation was previously derived in 1D homogeneous gases  \cite{Buchhold2015, Buchhold2016}. In this case, $n_{x,k,t}$ becomes independent of $x$ and essentially reduces to the phonon momentum distribution. For an inhomogeneous gas, in contrast, the kinetic equation depends explicitly on $x$ and reads (see Sec. \ref{Sec:FieldTheory} for details):
\begin{widetext}
\begin{equation}
\label{eq:kinetic}
\begin{split}
    (\partial_t+c\,\partial_x)n_{x,k,t} = &-i\int_0^\infty\frac{\mathrm{d}p}{\pi\rho_0}~\frac{4\Lambda_{k,p}^2}{\Sigma^R_{x,k+p,t}-\Sigma^{R*}_{x,p,t}-\Sigma^{R*}_{x,k,t}}[n_{x,p+k,t}(n_{x,p,t}+n_{x,k,t}+1)-n_{x,p,t}n_{x,k,t}] \\
    &-i\int_0^k\frac{\mathrm{d}p}{\pi\rho_0}~\frac{2\Lambda_{p,k-p}^2}{\Sigma^R_{x,p,t}+\Sigma^R_{x,k-p,t}-\Sigma^{R*}_{x,k,t}}[n_{x,p,t}n_{x,k-p,t}-n_{x,k,t}(n_{x,p,t}+n_{x,k-p,t}+1)],
\end{split}
\end{equation}
\end{widetext}
where $\Lambda_{k,p}$ has been defined below Eq.~\eqref{eq:Hhydroa}. The left-hand side of Eq.~\eqref{eq:kinetic} describes the propagation of sound-wave excitations in the superfluid, while the right-hand side accounts for their damping due to phonon-phonon interactions. The two collision integrals represent distinct types of scattering processes: Landau processes (first line), corresponding to the recombination of two phonons, $(p,k)\to p+k$, and Beliaev processes (second line), corresponding to the decay of a phonon into two others, $k\to(p,k-p)$.

\subsection{Long-time equilibrium}

Before looking at the relaxation dynamics, let us first examine the long-time solution of the problem, when the system has rethermalized. When $t\to\infty$, the initial spatial modulation is completely damped so all observables become independent of $x$. In particular, the Wigner distribution converges to the stationary and homogeneous solution of Eq.~\eqref{eq:kinetic} that simultaneously nullifies both collision integrals. This solution is
the Bose-Einstein distribution
\begin{equation}\begin{aligned}
\label{eq:BEinfinite}
    n_{x,k,\infty}\equiv n_{k}=\frac{1}{\exp(\epsilon_k/T_f)-1},
\end{aligned}\end{equation}
where $T_f$ is the final temperature reached by the system long after the quench. The value of $T_f$ is fixed by the condition of energy conservation, $\int_k c|k| n_{k,0}=\int_k c|k| n_{k}$, where $n_{k,0}$ is given by Eq.~\eqref{eq:pq_mom} \cite{footnote}. This leads to
\begin{equation}
    T_f = T_0\sqrt{1+2\varepsilon^2\left[1+\frac{2\pi^2}{3}\left(\frac{T_0}{k_0c}\right)^2\right]}
    \label{eq:Tf}
\end{equation} 
which is, as expected, above $T_0$ since the initial periodic modulation injects energy into the system.

The  phonon scattering rate in this final equilibrium state, $\smash{\gamma_k\equiv i\Sigma^R_{x,k,\infty}}>0$ (recall that $\smash{\Sigma^R_{x,k,t}}$ is purely imaginary), is then obtained from Eq.~\eqref{eq:self-energyR} by substituting $n_{x,k,t}\to n_{k}$ and $i\Sigma^R_{x,k,t}\to\gamma_k$. This yields
\begin{equation}\begin{aligned}
    \gamma_k =& \phantom{.}2\int_0^\infty\frac{\mathrm{d}p}{\pi\rho_0}~\frac{\Lambda_{k,p}^2(n_{p}-n_{k+p})}{\gamma_{k+p}+\gamma_p+\gamma_k} \phantom{\Bigg[_{[}}\\ 
    &+\int_0^k\frac{\mathrm{d}p}{\pi\rho_0}~\frac{\Lambda_{p,k-p}^2(n_{k-p}+n_{p}+1)}{\gamma_{k-p}+\gamma_p+\gamma_k}.
    \label{eq:gammaeq}
\end{aligned}\end{equation}
This expression encapsulates two different regimes for phonon scattering. The first one, the Beliaev regime, corresponds to momenta $k\gg T_f/c$, such that $n_{k}\ll1$. Equation~\eqref{eq:gammaeq} then simplifies to 
\begin{equation}
\label{eq:Beliaev}
    \gamma_k^B\simeq  \int_0^k\frac{\mathrm{d}p}{\pi\rho_0}~\frac{\Lambda_{p,k-p}^2}{\gamma_{k-p}+\gamma_p+\gamma_k},
\end{equation}
which leads to the 1D Beliaev scattering rate \cite{Punk2006} 
\begin{equation}
    \gamma_{k}^B= \frac{3(ck)^2}{8g\rho_0}\sqrt{\frac{2\pi/(3\sqrt{3})-1}{2\pi\rho_0\xi}}._{\phantom{\Big(}}
\end{equation}
In the opposite Landau regime where we have $k\ll T_f/c$, $n_{k}\simeq T_f/(c|k|)\gg1$, and defining $g_{L} =  \left( \frac{3k}{4mc} \right)^{2} \frac{gT_f}{2\pi}$ we find that  Eq.~\eqref{eq:gammaeq} can be approximated by
\begin{equation}
\gamma_k^L\simeq {g_{L}} \!\left[\int_0^\infty \!\!\frac{\mathrm{d}p}{\gamma_{k+p}\!+\!\gamma_p\!+\!\gamma_k}\!+\!
\int_0^k\!\frac{\mathrm{d}p/2}{\gamma_{k-p}\!+\!\gamma_p\!+\!\gamma_k}\right]. \vspace{1em}
\end{equation}
This leads to the 1D Landau scattering rate:
\begin{equation}\label{eq:Landau}
    \gamma_k^L= \frac{3T_f^2}{8g\rho_0}\sqrt{\frac{2.74}{\pi\rho_0\xi}}\left(\frac{ck}{T_f}\right)^{3/2},
\end{equation}
which displays a characteristic nonanalytic scaling in $k^{3/2}$ at low momenta. This scaling  was previously confirmed in truncated-Wigner numerical simulations \cite{Kulkarni2013}.

\subsection{Near-equilibrium relaxation}

We now return to the quench dynamics. From Sec. \ref{Sec:density_relax}, it follows that a full determination of the density $\rho(x,t)$ in principle requires simultaneously solving the local coupled equations \eqref{eq:self-energyR} and \eqref{eq:kinetic} for $\Sigma^R_{x,k,t}$ and $n_{x,k,t}$.
This procedure is numerically demanding; however, it can be significantly simplified in the case of a weak quench considered here. Indeed, since $\epsilon\ll1$, the gas remains at any time close to its long-time equilibrium state described by the thermal distribution \eqref{eq:BEinfinite} and by the phonon scattering rate \eqref{eq:gammaeq}. This invites us to write \cite{footnote2}
\begin{equation}\begin{aligned}
\label{eq:perturbationnsigma}
    n_{x,k,t}\simeq n_{k}+\delta n_{x,k,t},\ \ 
    i\Sigma^R_{x,k,t}\simeq \gamma_k+i\delta\Sigma^R_{x,k,t},   
\end{aligned}\end{equation}
where $\delta n_{x,k,t}\ll n_{k}$ and $|\delta\Sigma^R_{x,k,t}|\ll\gamma_k$.
Inserting these expansions into Eqs. \eqref{eq:self-energyR} and \eqref{eq:kinetic}, linearizing and integrating over the position $x$, we obtain the following set of \emph{global} coupled equations:
\begin{widetext}
\begin{equation}
\label{eq:deltagamma}
\begin{split}
    \delta\gamma_{k,t} = 2&\int_0^\infty\frac{\mathrm{d}p}{\pi\rho_0}~\frac{\Lambda_{k,p}^2}{\gamma_{k+p}+\gamma_p+\gamma_k}[(\delta n_{p,t}-\delta n_{k+p,t})-\frac{n_p-n_{k+p}}{\gamma_{k+p}+\gamma_p+\gamma_k}(\delta\gamma_{k+p,t}+\delta\gamma_{p,t}+\delta\gamma_{k,t})]\\
    +&\int_0^k\frac{\mathrm{d}p}{\pi\rho_0}~\frac{\Lambda_{p,k-p}^2}{\gamma_{p}+\gamma_{k-p}+\gamma_k}[(\delta n_{k-p,t}+\delta n_{p,t})-\frac{n_{k-p}+n_{p}+1}{\gamma_p+\gamma_{k-p}+\gamma_k}(\delta\gamma_{p,t}+\delta\gamma_{k-p,t}+\delta\gamma_{k,t})]
\end{split}
\end{equation}
and
\begin{equation}
\label{eq:deltan}
\begin{split}
    \partial_t\delta n_{k,t} = &\int_0^\infty\frac{\mathrm{d}p}{\pi\rho_0}~\frac{4\Lambda_{k,p}^2}{\gamma_{k+p}+\gamma_{p}+\gamma_{k}}\left[\delta n_{k,t}(n_{p+k}-n_{p})+\delta n_{p,t}(n_{p+k}-n_{k})+\delta n_{p+k,t}(n_{p}+n_{k}+1)\right] \phantom{\Bigg[_{[]}} \\
    +&\int_0^k\frac{\mathrm{d}p}{\pi\rho_0}~\frac{2\Lambda_{p,k-p}^2}{\gamma_{p}+\gamma_{k-p}+\gamma_{k}}\left[\delta n_{p,t}(n_{k-p}-n_{k})+\delta n_{k-p,t}(n_{p}-n_{k})-\delta n_{k,t}(n_{p}+n_{k-p}+1)\right],
\end{split}
\end{equation}
\end{widetext}
where we defined $\delta\gamma_{k,t}\equiv i\int_x\delta\Sigma^R_{x,k,t}$, the nonequilibrium deviation to $\gamma_k$, and $\delta n_{k,t}\equiv \int_x\delta n_{x,k,t}$, the nonequilibrium deviation to $n_k$. Compared to a direct computation of Eqs. \eqref{eq:self-energyR} and \eqref{eq:kinetic}, the task is now considerably simplified since the $x$ dependence has been integrated out. Resolution of Eqs. \eqref{eq:deltagamma} and \eqref{eq:deltan} is done iteratively, using a single initial condition, $\delta n_{k,0}$, given by the second term on the right-hand side of Eq.~\eqref{eq:pq_mom}.

Once $\delta\gamma_{k,t}$ is determined, the density of the Bose gas follows from Eqs. \eqref{eq:density_a} and \eqref{eq:Akt}:
\begin{align}
\label{eq:rhofinal}
    \rho(x,t)=
    \rho_0\Big[1-&\frac{V_0 \epsilon_{k_0}\delta t}{g\rho_0}\cos(k_0x)
    \sin[\epsilon_{k_0}(t+\delta t/2)]\nonumber\\
    &\times
    \exp\big(-\gamma_{k_0}t
    -\int_0^tdt'\,\delta\gamma_{k_0,t'}\big)\Big].
\end{align}
At sufficiently long time, the density relaxation exhibits the expected exponential behavior, governed by the \emph{equilibrium} phonon scattering rate $\gamma_{k_0}$. However, at earlier times deviations from this exponential form arise through the factor $\smash{\int_0^tdt'\,\delta\gamma_{k_0,t'}}$. These non-exponential corrections are the focus of the next section.

\section{Results}
\label{Sec:results}

\subsection{Phonon distribution}

\begin{figure}
\centering
\includegraphics[width=8cm]{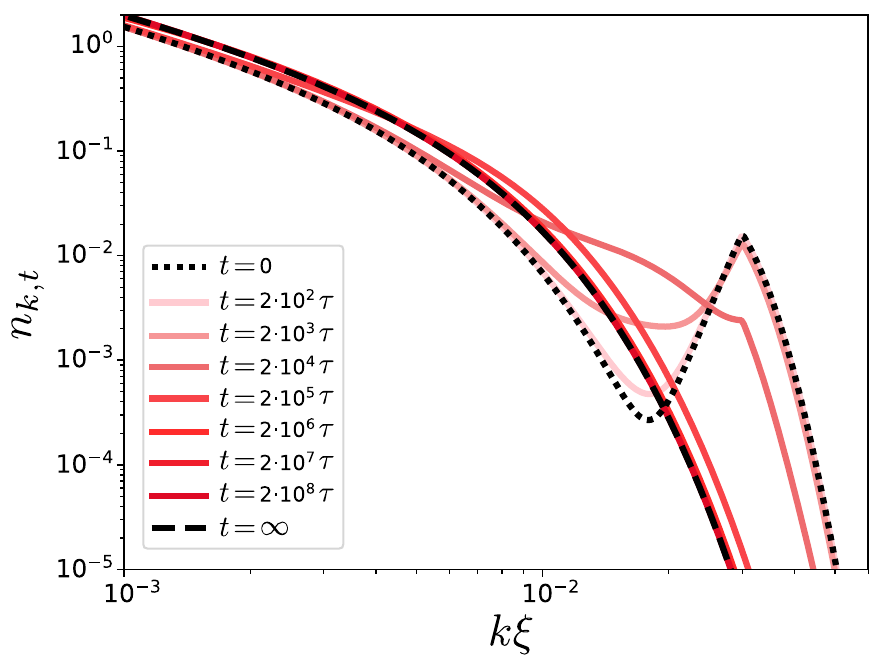}
\caption{\label{Fig_momt}
Phonon distribution $n_{k,t}$ at successive times following the periodic-potential quench, computed numerically from Eq.~\eqref{eq:deltan}. Time is expressed in units of $\tau\equiv 1/(2g\rho_0)$.
At $t=0$, the distribution features a triangular peak centered at $k=k_0$, superimposed on a smooth thermal background at temperature $T_0$, Eq.~\eqref{eq:pq_mom}. In the long-time limit $t=\infty$, relaxation is complete and $n_{k,\infty}$ coincides with the thermal distribution \eqref{eq:BEinfinite} at $T_f>T_0$. Parameters are chosen as $T_0/(2g\rho_0)=2.10^{-3}$, $\rho_0\xi=5$ (weakly interacting gas), $k_0\xi=3.10^{-2}$ and $T_f = T_0\sqrt{1+0.5}$. $\varepsilon$ is determined by inverting Eq. \eqref{eq:Tf}, yielding $\varepsilon^2 = 0.24$.}
\end{figure}

\begin{figure}
\centering
\includegraphics[width=8cm]{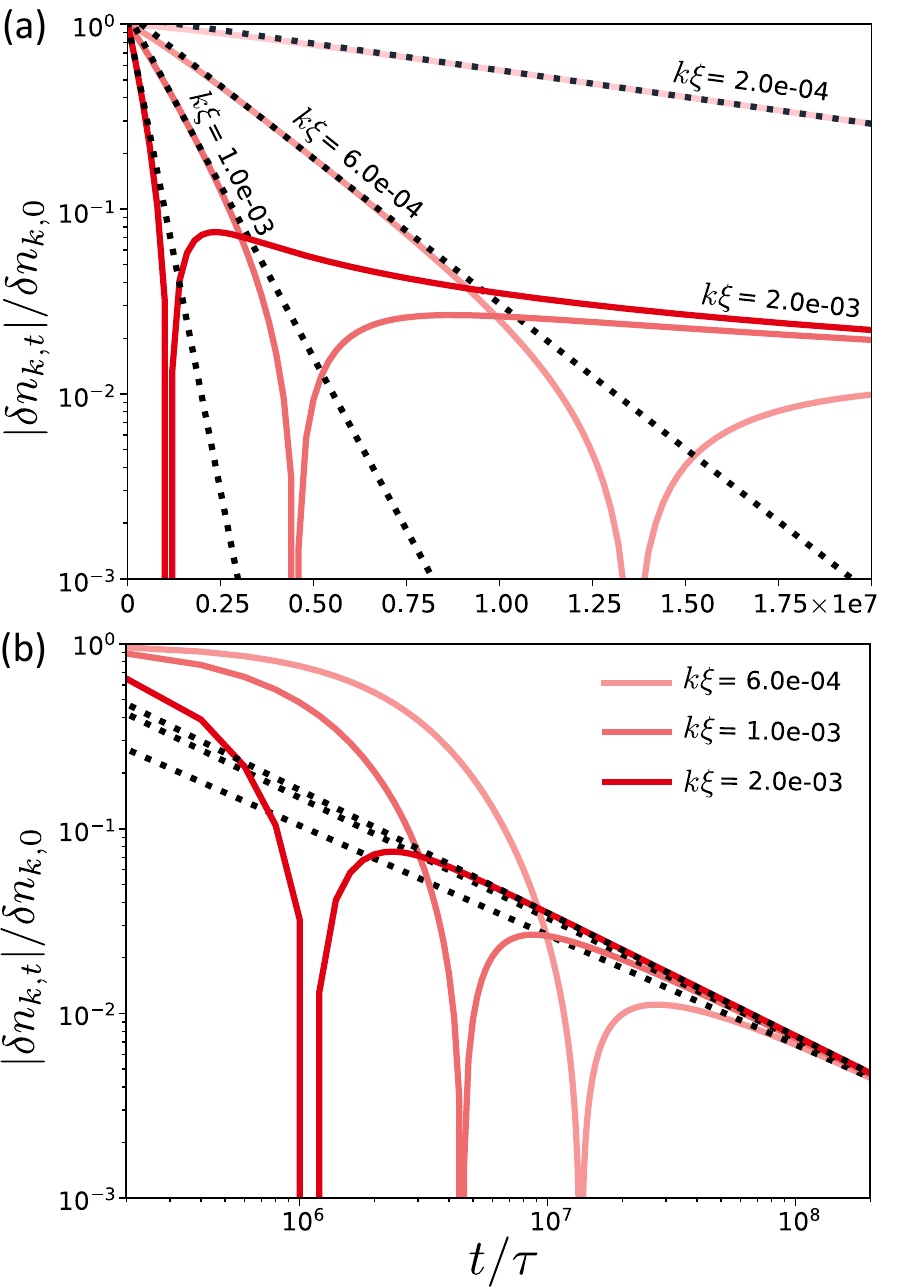}
\caption{\label{Fig_mom_expalg}
Deviation $|\delta n_{k,t}|\equiv |n_{k,t}-n_k|$ of the phonon distribution from its equilibrium value as a function of time, shown in (a) logarithmic  and (b) log-log scales, for several momentum values (dips in the curves correspond to a change of sign of $\delta n_{k,t}$). 
The deviation decays exponentially at short times and algebraically at long times. 
Solid curves correspond to numerical solutions of Eq.~\eqref{eq:deltan}. 
In (a), dotted lines are exponential fits of the form $\exp(-1.5\gamma_k t),\, \exp(-1.7\gamma_k t),\ \exp(-1.9\gamma_k t)$ and $\exp(-2.1\gamma_k t)$ from top to bottom.
In (b), dotted lines are algebraic fits of the form $t^{-\alpha}$, with $\alpha=0.593, 0.651$ and $0.666$ from smallest to highest $k\xi$ values. Parameters used in the numerics are the same as in Fig. \ref{Fig_momt}.}
\end{figure}

We first show in Fig. \ref{Fig_momt} the phonon momentum distribution $n_{k,t}$ at different times, obtained by numerically solving Eqs.~\eqref{eq:deltagamma} and \eqref{eq:deltan} starting from the post-quench initial condition \eqref{eq:pq_mom}. At $t=0$, the distribution exhibits a pronounced peak around $k=k_0$, originating from the periodic-potential quench. As the system evolves, this peak gradually smoothens and eventually disappears, while $n_{k,t}$ slowly converges to the thermal distribution~\eqref{eq:BEinfinite}. 
To examine more closely the approach to equilibrium, we show in Fig. \ref{Fig_mom_expalg} the deviation of the phonon distribution from its equilibrium profile $\delta n_{k,t}=n_{k,t}-n_{k}$.
Figure \ref{Fig_mom_expalg}(a) displays this deviation in log scale, and Fig.~\ref{Fig_mom_expalg}(b) in log-log scale. These representations show that the decay of $\delta n_k$ is initially exponential, and becomes algebraic at late times. 
This cross-over from exponential to algebraic relaxation was identified in previous works in 1D \cite{Lin2013, Lux2014, Buchhold2015} and 2D \cite{Duval2025} Bose gases. Within the linearized kinetic equation \eqref{eq:deltan}, the exponential decay originates from the diagonal terms (proportional to $\delta n_{k,t}$) on the right-hand side. Indeed, neglecting the off-diagonal contributions in \eqref{eq:deltan} and comparing with the general expression \eqref{eq:gammaeq} for the equilibrium scattering rate $\gamma_k$, we readily obtain $\partial_t \delta n_{k,t}=-2\gamma_k \delta n_{k,t}$,  which leads to the exponential law $\delta n_{k,t}\propto\exp(-2\gamma_k t)$. 
Exponential fits to the numerical data shown in Fig. \ref{Fig_mom_expalg}(a) are in close agreement with this prediction. 
At longer times, however, the relaxation becomes algebraic. This regime originates from the off-diagonal terms in Eq.~\eqref{eq:deltan}, which dominate at late times. 
These contributions are supported by slow modes constrained by energy conservation \cite{Buchhold2015, Duval2025}. They lead to  $\delta n_{k,t}\propto t^{-\alpha}$, with $\alpha\simeq 2/3$. The physical origin of this exponent can be qualitatively understood from the following argument, first introduced in \cite{Duval2025} for 2D Bose gases. 
The condition of energy conservation is 
\begin{equation}
\label{eq:energy_conservation}
    \int_kc\,k\,\delta n_k(t) = 0,
\end{equation}
and holds at all times. To analyze its implications, we divide the momentum integral into three regions: $0<k<k_e(t)$,  $k_e(t)<k<k_a(t)$ and $k_a(t)<k$, corresponding, respectively, to (i) the ``prethermal'' regime where phonon scattering is negligible, i.e., $\delta n_{k,t}\simeq \delta n_{k,0}$; (ii) an intermediate regime where relaxation is exponential, $\delta n_{k,t}\simeq \delta n_{k,0}e^{-2\gamma_kt}$, and (iii) a high-momentum regime characterized by algebraic decay, $\delta n_{k,t}\simeq f(k)/t^\alpha$. With this simple model, Eq.~\eqref{eq:energy_conservation} can  be rewritten as
\begin{equation}
\label{eq:EC_split}
   - \int_{k_a}^\infty\!\frac{\mathrm{d}kkf(k)}{t^\alpha} \!\simeq\!\int_{0}^{k_e}\!\! \mathrm{d}k k\delta n_{k,0}\!+\! \int_{k_e}^{k_a}\!\! \mathrm{d}k k\delta n_{k,0}e^{-2\gamma_kt}.
\end{equation}
The crossover scale $k_e(t)$ is defined by the condition $e^{-2\gamma_{k_e}t}\sim 1$. In the Landau regime where $\gamma_k$ is given by Eq.~\eqref{eq:Landau}, this leads to $k_e(t)\sim 1/t^{2/3}$. The second crossover scale, $k_a(t)$, is defined by $e^{-2\gamma_{k_a}t}\sim f(k_a)/t^\alpha$. Assuming that the function $f$ is smooth at low momenta, one finds $k_a(t)\sim(\ln t/t)^{2/3}$. The post-quench distribution, on the other hand, scales as $\delta n_{k,0}\sim 1/k$ at low momenta [see Eq.~\eqref{eq:pq_mom}]. Inserting these results in the right-hand side of Eq.~\eqref{eq:EC_split}, we finally obtain 
\begin{equation}
   - \int_{k_a}^\infty\!\frac{\mathrm{d}kkf(k)}{t^\alpha} \underset{t\to\infty}{\longrightarrow}
   \int_{0}^\infty\!\frac{\mathrm{d}kkf(k)}{t^\alpha}\!\sim\!\frac{1}{t^{2/3}},
\end{equation}
which demonstrates that $\alpha=2/3$. Algebraic fits to the numerical results shown in Fig. \ref{Fig_mom_expalg}(b) are consistent with this prediction. The agreement is better at larger $k\xi$ values, where the algebraic regime sets in earlier, providing a longer time window for the fit.

\subsection{Nonequilibrium scattering rate}
\begin{figure}[h]
\centering
\includegraphics[width=8cm]{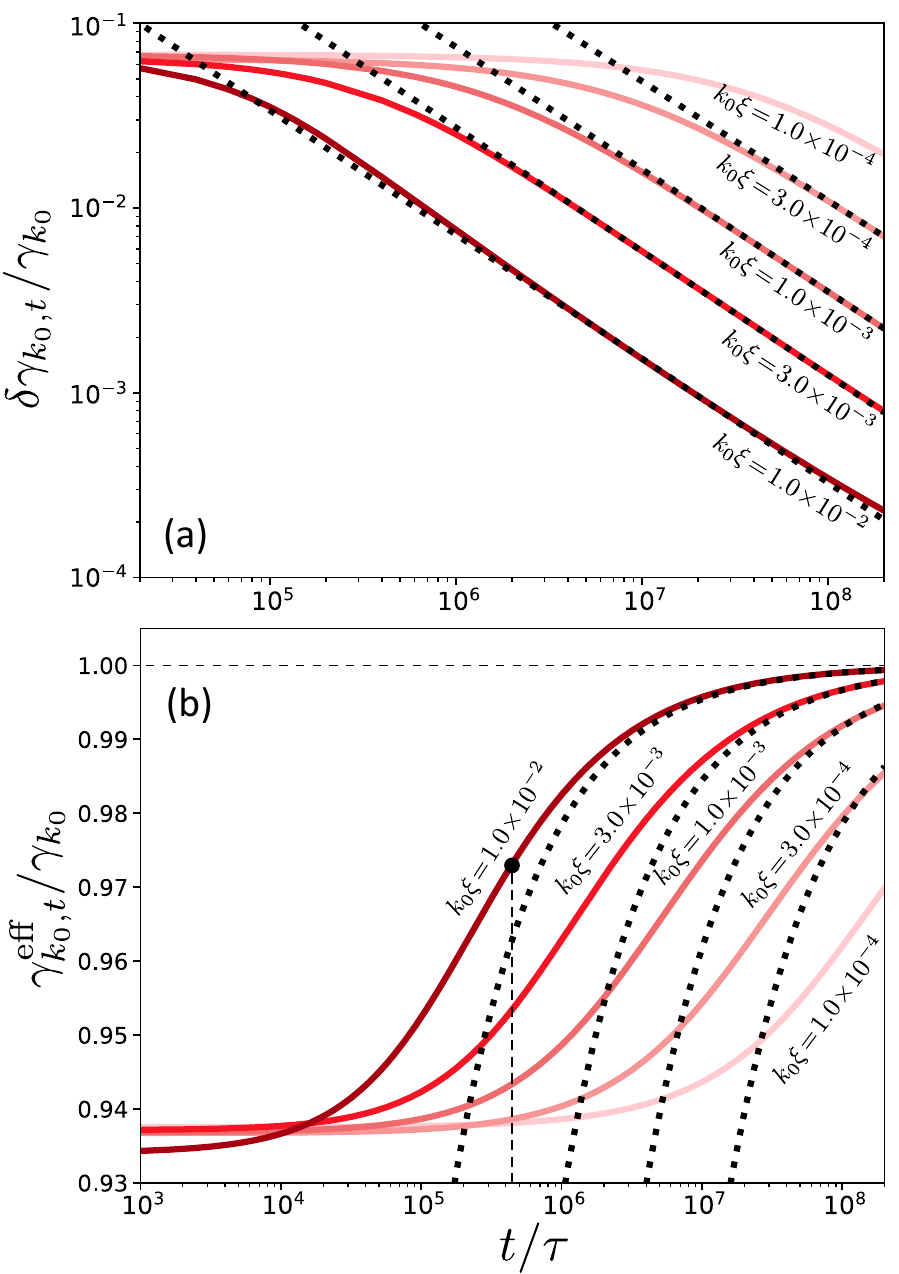}
\caption{\label{FIg_SEDecay}
(a) Nonequilibrium deviation $\delta\gamma_{k_0,t}$ of the  scattering rate from its thermal value $\gamma_{k_0}$ as a function of time, shown for several probe momenta $k_0$. Solid lines correspond to numerical results from Eq.~\eqref{eq:deltagamma}, and dotted lines are power-law fits $\propto t^{-\alpha}$,  with $\alpha=0.646, 0.662, 0.669$ and $0.671$ from top to bottom. 
(b) Effective scattering rate, Eq.~\eqref{eq:gammakeff}, as a function of time, for various $k_0$. At long time, $\gamma^\text{eff}_{k_0,t}\to \gamma_{k_0}$. Dotted curves are fits to a function of the form $1-\text{const}/t^\alpha$, using the same $\alpha$ values as in (a). The black dot marks the inflection point defining the relaxation time $t_\Sigma$. 
The final temperature is set to $T_f=T_0\sqrt{1+0.5}$ for all curves, which corresponds to $\varepsilon^2 = 9.10^{-5}, 9.10^{-4}, 9.10^{-3}, 6.10^{-2}$ and $2.10^{-1}$ from smallest to highest $k_0\xi$ values. Parameters used in
the numerics are the same as in Fig. \ref{Fig_momt}.}
\end{figure}
We now turn to the numerical analysis of the density relaxation, Eq.~\eqref{eq:rhofinal}, which constitutes the main objective of this work. 
This relaxation is governed by
the nonequilibrium deviation $\delta\gamma_{k_0,t}$ from the equilibrium scattering rate $\gamma_{k_0}$, which is shown in Fig. \ref{FIg_SEDecay}(a) for several values of the probe momentum $k_0$. 
The numerical procedure to obtain these curves is as follows. We first fix a value of $k_0$, which determines the initial condition \eqref{eq:pq_mom}. 
We then compute the phonon distribution $n_{k,t}$ using Eq.~\eqref{eq:deltan}. Finally, the resulting distribution is inserted into Eq.~\eqref{eq:deltagamma} to evaluate $\delta\gamma_{k_0,t}$, which is then plotted as a function of time for $k=k_0$. The entire procedure is repeated for different $k_0$ values.

The plot in Fig. \ref{FIg_SEDecay}(a) shows that, after a relatively short transient, $\delta\gamma_{k,t}$ decays almost purely algebraically with no visible trace of exponential relaxation. This behavior can be understood by a simple inspection of \eqref{eq:deltagamma}: the time evolution of $\delta\gamma_{k,t}$ is governed solely by the off-diagonal contributions $\delta n_{p\ne k,t}$ in the phonon distribution. Any exponential relaxation present in $\delta n_{p\ne k,t}$ is smoothed out by the integration over $p$, leaving only the long-time algebraic decay of the phonon distribution to govern $\delta\gamma_k$. Since $\delta n_{k,t}\sim t^{-2/3}$, we immediately deduce
\begin{equation}
    \delta \gamma_{k_0,t}\sim t^{-2/3}
\end{equation}
This behavior is confirmed by algebraic fits to numerical data shown in Fig. \ref{FIg_SEDecay}(a).

Complementary to the instantaneous scattering rate $\gamma_{k,t}$, it is also useful to define an effective, time-dependent phonon scattering rate as
\begin{equation}
\label{eq:gammakeff}
    \gamma^\text{eff}_{k_0,t}\equiv \frac{1}{t}\int_0^t dt'\gamma_{k_0,t'}=\gamma_{k}+\frac{1}{t}\int_0^t dt'\delta \gamma_{k_0,t'},
\end{equation}
such that the density relaxes according to $\rho(x,t)-\rho_0\propto \exp(-\gamma^\text{eff}_{k_0,t}t)$. 
This quantity is shown in Fig.~\ref{FIg_SEDecay}(b) as a function of time.
Starting from its initial post-quench value --- which depends on the initial condition --- we find that $\smash{\gamma^\text{eff}_{k_0,t}}$ slowly decays toward its long-time equilibrium value $\gamma_{k_0}$ following the scaling law $\smash{\gamma^\text{eff}_{k_0,t}\simeq \gamma_{k_0}-\text{const}\times t^{-2/3}}$. Long-time fits to this law, shown in \ref{FIg_SEDecay}(b), provide an excellent description of the numerical results.

\subsection{Scattering rate relaxation time}

\begin{figure}[h]
\centering
\includegraphics[width=8cm]{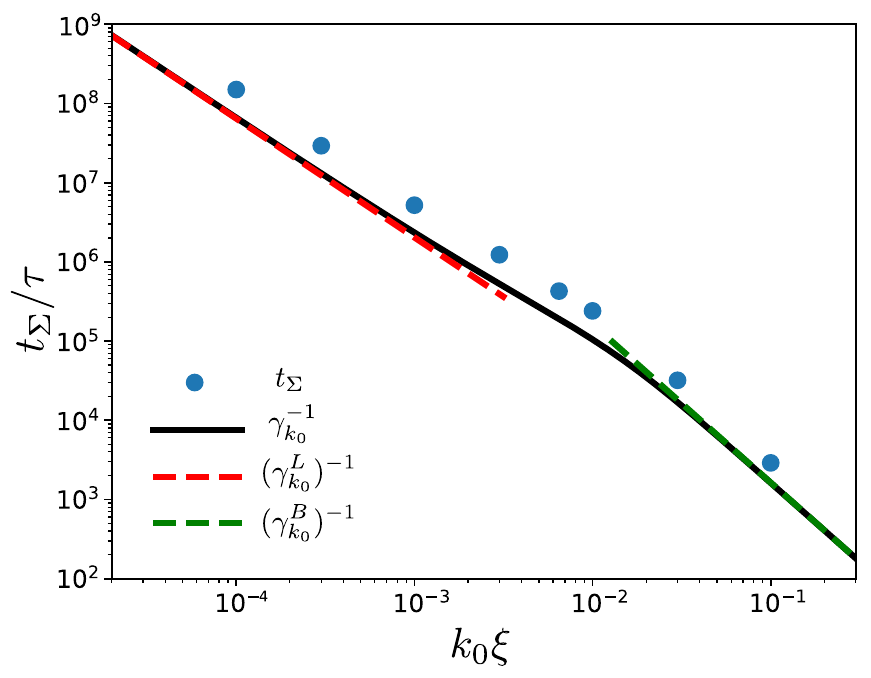}
\caption{\label{Fig:Thermaltime}
Relaxation time $t_\Sigma$ of the scattering rate $\gamma_{k_0,t}$ (blue dots), defined from the inflection point of the $\smash{\gamma_{k_0,t}^\text{eff}}$ curves in Fig. \ref{FIg_SEDecay}(b). The black solid curve shows, for comparison, the inverse of the equilibrium scattering rate, $\gamma_k^{-1}$, while the red (green) dashed line indicates its Landau (Beliaev) limit, given by Eq.  \eqref{eq:Landau} [Eq.\eqref{eq:Beliaev}]. Parameters used in
the numerics are the same as in Fig. \ref{Fig_momt}.
}
\end{figure}

We finally examine the characteristic time  $t_\Sigma$ required for the time-dependent scattering rate $\gamma_{k_0,t}$ to reach its equilibrium value $\gamma_{k_0}$. This timescale is identified from the inflection point of the $\smash{\gamma_{k_0,t}^\text{eff}}$ curves in Fig. \ref{FIg_SEDecay}(b). The resulting values of $t_\Sigma$ are plotted in Fig. \ref{Fig:Thermaltime} as a function of the probe momentum $k_0$ (blue dots).
For comparison, we also display on the same graph the  inverse of the equilibrium scattering rate $\gamma_k^{-1}$, obtained by numerically solving Eq.~\eqref{eq:gammaeq}, together with its asymptotic limits $\gamma^L_{k_0}$ and $\gamma^B_{k_0}$, Eqs. \eqref{eq:Landau} and \eqref{eq:Beliaev}. This shows that the relaxation time associated with $\gamma_{k_0,t}$ is itself of the order of the equilibrium phonon lifetime.

\section{Keldysh theory of the inhomogeneous Bose gas}
\label{Sec:FieldTheory}

In this section, we derive the dynamical equations \eqref{eq:Akt}, \eqref{eq:self-energyR}, and \eqref{eq:kinetic}. 
This amounts to obtaining a closed set of equations of motion for the coupled dynamics of $\langle\hat a_{k,t}\rangle$  and $n_{k,t}$, in the spirit of the traditional treatment of three-dimensional Bose–Einstein condensates \cite{Griffin2009, Kamenev2011}, where fluctuations evolve on top of a finite mean field—except that here the evolution takes place in the quasiparticle basis. To this end, we employ the powerful  framework of closed-time contours, which forms the cornerstone of nonequilibrium quantum field theory \cite{Keldysh1965, Altland2010, Kamenev2011, Rammer2007}.

We begin by defining the Keldysh generating functional, from which the one-particle-irreducible (1PI) effective action and the corresponding self-energies can be consistently obtained. The kinetic equations then follow perturbatively from this effective action.

\subsection{Generating functionals}

In the Keldysh field formalism, quantum mechanical averages are computed by a path-integral over a closed time contour \cite{Keldysh1965, Altland2010, Kamenev2011, Rammer2007}. This contour  is made of two branches, running from $t=0$ to $t=\infty$ and back, respectively, which results in a doubling of the degrees of freedom in the corresponding field theory. The Keldysh partition function is then defined as
\begin{equation}
    Z = \int\mathcal{D}[a_+,a_+^*,a_-,a_-^*]e^{iS[a_+,a_+^*]-iS[a_-,a_-^*]}
\end{equation}
where $S=S_0+S_{\text{int}}$ denotes the nonequilibrium hydrodynamic action in the coherent state representation [obtained from the Hamiltonian \eqref{eq:Hhydroa}], and $a_\pm$ are the scalar phononic fields defined on each branch \cite{Duval2023, Buchhold2015}. The first step is to perform a Keldysh rotation by introducing the ``classical'' and ``quantum'' fields $\alpha = (a_++a_-)/\sqrt{2}$ and $\tilde{\alpha} = (a_+-a_-)/\sqrt{2}$, which allows us to rewrite the contour integral in the action as a single integral over positive times. The quadratic part of the action then becomes
\begin{equation}
    S_0 = \int_{k,t>0}~ \left( \begin{array}{cc} \alpha_{k,t}^* & \tilde{\alpha}_{k,t}^* \end{array} \right) [G^0]^{-1}_{k,t} \left( \begin{array}{c} \alpha_{k,t} \\ \tilde{\alpha}_{k,t} \end{array} \right)\text{,}
    \label{eq:S0}
\end{equation}
with $\int_{t>0} \equiv \int_0^\infty dt$ and
\begin{equation}
    [G^0]^{-1}_{k,t} = \left( \begin{array}{cc} 0 & i\partial_t-\epsilon_k-i0^+ \\ i\partial_t-\epsilon_k+i0^+ & \phantom{.}2i0^+(2n_{k,t}+1) \end{array} \right)
\end{equation}
where we introduced the  $0^+$ as regularizations of the poles. The interaction part of the action reads 
\begin{equation}
\begin{split}
    S_{\text{int}} =&~\frac{1}{\sqrt{2}}\int_{k,p,t>0}~\Lambda_{k,p}(2\alpha_{k+p,t}^*\tilde{\alpha}_{p,t}\alpha_{k,t} \\
    &+\tilde{\alpha}_{k+p,t}^*\alpha_{p,t}\alpha_{k,t}+\tilde{\alpha}_{k+p,t}^*\tilde{\alpha}_{p,t}\tilde{\alpha}_{k,t}+\text{c.c.})\text{.}
    \label{eq:Sint}
\end{split}
\end{equation}
By introducing classical sources to the resulting partition function, one obtains the generating functional \cite{Rammer2007, Gasenzer2009, Sieberer2016, Stoof1999}
\begin{equation}
    Z[\{J\}] = \int\mathcal{D}[\{\alpha\}]e^{iS[\{\alpha\}]+i\int_{k,t>0}\sum_i\alpha^i_{k,t}J^i_{k,t}}
    \label{eq:Zexpression}
\end{equation}
where $\alpha^i \equiv \{\alpha, \alpha^*, \tilde{\alpha}, \tilde{\alpha}^*\}$ and $J^i \equiv \{\tilde{J}^*, \tilde{J}, J^*, J\}$. From the latter, one can generate $n$-point correlators by performing $n$ functional derivatives with respect to the classical sources $J^i$. 

The partition function $Z$ generates all correlation functions, including both connected and disconnected contributions.
To isolate the connected parts, we introduce the Schwinger functional $W[\{J\}] \equiv -i\ln{Z[\{J\}]}$ \cite{Rammer2007, Chou1984, Berges2007, Gasenzer2009, Sieberer2016}, from which the one- and two-point connected correlators follow as
\begin{equation}
\begin{aligned}
        \mathcal{A}^i_{k,t} \equiv& \ \langle\alpha^i_{k,t}\rangle = \frac{\delta W[\{J\}]}{\delta J^i_{k,t}} = \frac{-i}{Z[\{J\}]}\frac{\delta Z[\{J\}]}{\delta J^i_{k,t}}\\
    G^{ij}_{k,k',t,t'} \equiv& -i\langle\alpha^i_{k,t}\alpha^j_{k',t'}\rangle_c = -\frac{\delta^2W[\{J\}]}{\delta J^i_{k,t}\delta J^j_{k',t'}}\text{.}
    \label{eq:Adef}
\end{aligned}
\end{equation}
The one-point functions correspond to macroscopic fields that can be related to the expectation values of the phonon operators.
Owing to the definition of the Keldysh rotation, one finds $\mathcal{A}_k = \sqrt{2}\langle\hat{a}_k\rangle$ and $\tilde{\mathcal{A}}_k = 0$ in the absence of external sources. The two-point connected correlators, also referred to as propagators, 
can be collected into a $4\times4$  matrix defined in the so-called Keldysh–Nambu space \cite{Buchhold2015, Buchhold2016, Sieberer2016}. 
However, because of fundamental relations among its entries, only four of them are independent, one of which corresponds to an anomalous propagator that is irrelevant for the present discussion. For this reason, we restrict the full matrix to its Keldysh subspace, defined by
\begin{equation}
    G_{k,k',t,t'} \equiv \left( \begin{array}{cc} G^{K}_{k,k',t,t'} & G^{R}_{k,k',t,t'} \\ G^{A}_{k,k',t,t'} & 0 \end{array} \right),
    \label{eq:Gkktt}
\end{equation}
where $iG^R_{k,k',t,t'} = \langle\alpha_{k,t}\tilde{\alpha}^*_{k',t'}\rangle$, $iG^A_{k,k',t,t'} = \langle\tilde{\alpha}_{k,t}\alpha^*_{k',t'}\rangle$ and $iG^K_{k,k',t,t'} = \langle\alpha_{k,t}\alpha^*_{k',t'}\rangle$  denote the retarded, advanced and Keldysh Green's functions, respectively \cite{Duval2023, Buchhold2015, Sieberer2016}. The retarded and advanced components describe the system’s linear response to external perturbations, while the equal-time Keldysh Green’s function encodes information about its correlations, as 
\cite{Kamenev2011, Rammer2007}:
\begin{equation}
   i\int_{k'}G^K_{k+k'/2,k-k'/2,t,t}~e^{ik'x} = 2n_{x,k,t}+1,
\end{equation}
where $n_{x,k,t}$ is the phonon Wigner distribution (\ref{eq:wignerphonon}).

\subsection{1PI effective action}

The 1PI effective action is formally defined as the Legendre transform of $W$:
\begin{equation}
\label{eq:Gammadef}
    \Gamma[\{\mathcal{A}\}] \equiv W[\{J\}] - \int_{k,t>0}\sum_i\mathcal{A}^i_{k,t}J^i_{k,t},
\end{equation}
where the fields $\mathcal{A}^i$ are understood to depend implicitly on the sources $J^i$ \cite{Rammer2007, Gasenzer2009, Sieberer2016, Chou1984}. The 1PI effective action serves two main purposes: it provides the equations of motion for the macroscopic fields and allows one to compute the two-point correlator. The equation of motion for  $\mathcal{A}^i_k$ follows directly from the first line in Eq. (\ref{eq:Adef}), together with Eq. (\ref{eq:Gammadef}):
\begin{equation}
    \frac{\delta\Gamma[\{\mathcal{A}\}]}{\delta\mathcal{A}^i_{k,t}} = -J^i_{k,t}\text{.}
    \label{eq:HamiltonPrinciple}
\end{equation}
The two-point correlator, on the other hand, is obtained from the second functional derivatives of $\Gamma$ \cite{Rammer2007, Gasenzer2009, Sieberer2016, Chou1984}:
\begin{equation}
   \frac{\delta^2\Gamma[\{\mathcal{A}\}]}{\delta\mathcal{A}^i_{k,t}\delta\mathcal{A}^j_{k',t'}}= [G^{-1}]^{ij}_{k,k',t,t'} \text{.}
    \label{eq:gm1gamma}
\end{equation}

To determine the effective action $\Gamma$, we note that, using Eq. (\ref{eq:Gammadef}) together with the definition of $W$,  the $Z$ generating functional can be rewritten as
\begin{equation}
    Z[\{J\}] = e^{i\Gamma[\{\mathcal{A}\}]+i\int_{k,t>0}\sum_i\mathcal{A}^i_{k,t}J^i_{k,t}}.
    \label{eq:ZGamma}
\end{equation}
To evaluate $\Gamma$, we shift the integration variables in Eq. (\ref{eq:Zexpression}) as $\alpha^i \rightarrow \mathcal{A}^i+\alpha^i$, so that $\alpha^i$ now represent fluctuations around the macroscopic fields and vanish upon averaging \cite{Rammer2007, Berges2007} (``background field method''). Comparing Eqs.~(\ref{eq:Zexpression}) and (\ref{eq:ZGamma}) then yields the effective action in terms of an integral over the fluctuating fields:
\begin{equation}
\label{eq:eigamma}
    e^{i\Gamma[\{\mathcal{A}\}]} = \int\mathcal{D}[\{\alpha\}]e^{iS[\{\mathcal{A}\},\{\alpha\}]+i\int_{k,t>0}\sum_i\alpha^i_{k,t}J^i_{k,t}}\text{.}
\end{equation}
Expressed in terms of fluctuations around the macroscopic field, the microscopic action takes the form
\begin{equation}
\begin{split}
    S[\{\mathcal{A}\},\{\alpha\}] =&~S_0^{\mathcal{A}}[\{\mathcal{A}\}] + S_{\text{int}}^{\mathcal{A}}[\{\mathcal{A}\}] + S_0^\alpha[\{\alpha\}]_{\phantom{\big(}} \\
    &+ S_{\text{int}}^\alpha[\{\alpha\}] + S_2[\{\mathcal{A}\},\{\alpha\}],
    \label{eq:Sshift}
\end{split}
\end{equation}
where  we discarded linear terms in the fluctuating fields, which do not contribute to the 1PI effective action.
In Eq. (\ref{eq:Sshift}), $S_0^{\mathcal{A}}+S_{\text{int}}^{\mathcal{A}}$ and $S_0^\alpha+S_{\text{int}}^\alpha$ have the same structure as the original action given in \eqref{eq:S0} and \eqref{eq:Sint}, but with $\mathcal{A}^i$ or $\alpha^i$ as their respective field variables. The term coupling macroscopic and fluctuating components reads
\begin{align}
    S_2 =&~\frac{1}{\sqrt{2}}\int_{k,p,t>0}~\Lambda_{k,p}\Big[\tilde{\mathcal{A}}_{k+p,t}^*(\alpha_{p,t}\alpha_{k,t}+\tilde{\alpha}_{p,t}\tilde{\alpha}_{k,t}) \nonumber\\
    &+2\tilde{\mathcal{A}}_{k,t}^*(\alpha_{p,t}^*\alpha_{k+p,t}+\tilde{\alpha}_{p,t}^*\tilde{\alpha}_{k+p,t})+2\tilde{\alpha}_{p,t}^*\alpha_{k,t}^*\mathcal{A}_{k+p,t }\nonumber \\   &+2(\alpha_{k+p,t}^*\tilde{\alpha}_{p,t}+\tilde{\alpha}_{k+p,t}^*\alpha_{p,t})\mathcal{A}_{k,t}+\text{c.c.}\Big]\text{.}
\end{align}
Inserting Eq. (\ref{eq:Sshift}) into Eq. (\ref{eq:eigamma}) and setting the sources to zero, we infer:
\begin{equation}
    e^{i\Gamma[\{\mathcal{A}\}]} = e^{iS_0^{\mathcal{A}}+iS_{\text{int}}^{\mathcal{A}}+i\Delta\Gamma},
\end{equation}
where 
\begin{equation}
\label{eq:eigammadef}
    e^{i\Delta\Gamma}=
    \langle e^{iS_2+iS_{\text{int}}^\alpha}\rangle_0
\end{equation}
and the angular brackets $\langle\text{...}\rangle_0\equiv\int\mathcal{D}[\{\alpha\}]\text{...}\exp(iS_0^{\alpha})$ denote averaging over the free action of the fluctuating fields, $S_0^\alpha$. 
Expressing the effective action $\Gamma = S_0^{\mathcal{A}}+\Gamma_{\text{int}}$ as a sum of a free, $S_0^{\mathcal{A}}$, and an interacting component, $\Gamma_{\text{int}} \equiv S_{\text{int}}^{\mathcal{A}}+\Delta\Gamma$,
one recovers from Eq. (\ref{eq:gm1gamma}) the Dyson equation:
\begin{equation}
G^{-1} = G_0^{-1}-\Sigma,
\label{eq:Dyson}
\end{equation} 
with
\begin{equation}
    \Sigma^{ij}_{k,k',t,t'} \equiv -\frac{\delta^2\Gamma_{\text{int}}[\{\mathcal{A}\}]}{\delta\mathcal{A}^i_{k,t}\delta\mathcal{A}^j_{k',t'}}
    \label{eq:sigma}
\end{equation}
defining the self-energy.

\subsection{Equation of motion for the macroscopic field}

We now derive the equation of motion~\eqref{eq:Akt} for the macroscopic field $\mathcal{A}_{k,t}$ using a truncation scheme for the 1PI effective action.
To this end, we exploit the fact that $\mathcal{A}_k^i$ is small for a weak potential quench, typically of first order in $V_0/(g\rho_0)\ll 1$ [see Eq.~(\ref{eq:apostquench})]. This allows us to expand Eq.~(\ref{eq:eigammadef}) in powers of $S_2$, which is linear in the macroscopic field:
\begin{equation}
    e^{i\Delta\Gamma}~
    \simeq \langle e^{iS_{\text{int}}^\alpha}\rangle_0 + i\langle S_2e^{iS_{\text{int}}^\alpha}\rangle_0 - \frac{1}{2}\langle S_2^2e^{iS_{\text{int}}^\alpha}\rangle_0\text{.}
    \label{eq:TaylorExpansionDeltaGamma}
\end{equation}
The first term on the right-hand side, $\langle e^{iS_{\text{int}}^\alpha}\rangle_0$, vanishes identically, as it corresponds to the sum of source-independent vacuum bubble diagrams. These cancel due to the closed-time contour \cite{Kamenev2011, Rammer2007}. The second term, $\langle S_2 e^{iS_{\text{int}}^\alpha}\rangle_0$, also vanishes as a consequence of momentum conservation. To leading order, $\Delta\Gamma$ is therefore determined by the last term, which is quadratic in the macroscopic field. In particular, $\Delta\Gamma$ dominates over $S_{\text{int}}^{\mathcal{A}}$, which is cubic, so that $\Gamma_{\text{int}} \equiv S_{\text{int}}^{\mathcal{A}}+\Delta\Gamma\simeq \Delta\Gamma$. 
From the definition (\ref{eq:sigma}) of the self-energy, we infer that at leading order $\Delta \Gamma$ takes the form
\begin{align}
\label{eq:deltagamma_pert}
    \Delta\Gamma =&~-\int_{k,k',t,t'}[\Sigma^R_{k,k',t,t'}\mathcal{A}_{k,t}\tilde{\mathcal{A}}^*_{k',t'}+\Sigma^A_{k,k',t,t'}\tilde{\mathcal{A}_{k,t}}\mathcal{A}^*_{k',t'} \nonumber\\
    &+\Sigma^K_{k,k',t,t'}\tilde{\mathcal{A}}_{k,t}\tilde{\mathcal{A}}^*_{k',t'}+\text{anom.}] + o(\mathcal{A}_i^3),
\end{align}
where the self-energies are evaluated at ${\mathcal{A}}=0$. The ``anom.'' terms denote contributions proportional to
$\tilde{\mathcal{A}}_{k,t}\tilde{\mathcal{A}}_{k',t'}$ and $\tilde{\mathcal{A}}^*_{k,t}\tilde{\mathcal{A}}^*_{k',t'}$. They are irrelevant here, since they vanish upon taking $\smash{\tilde{\mathcal{A}}_{k,t}=0}$ at the end.

It is convenient to rewrite Eq.~(\ref{eq:deltagamma_pert}) using Wigner coordinates \cite{Kamenev2011, Rammer2007}. To do so, we redefine $(k+k')/2 \rightarrow k$ and $(t+t')/2 \to t$, and Fourier transform with respect to $k-k'$ and $t-t'$. Assuming further that the macroscopic fields vary slowly in space and time, $\Delta\Gamma$ becomes, in Wigner space,
\begin{align}
    \Delta\Gamma& =~-\int_{x,k,t,\omega}[\Sigma^R_{x,k,t,\omega}\mathcal{A}_{k,t}\tilde{\mathcal{A}}^*_{k,t}+\Sigma^A_{x,k,t,\omega}\tilde{\mathcal{A}}_{k,t}\mathcal{A}^*_{k,t} \nonumber\\
    &+\Sigma^K_{x,k,t,\omega}\tilde{\mathcal{A}}_{k,t}\tilde{\mathcal{A}}^*_{k,t}+\text{anom.}]B_{x,k,t,\omega} + o(\mathcal{A}_i^3)\text{,}
\end{align}
where 
$B_{x,k,t,\omega} \equiv i\big(G^R_{x,k,t,\omega}-G^A_{x,k,t,\omega}\big)$ is the spectral function, and we used the notation $\int_\omega \equiv \int_{-\infty}^{\infty} d\omega/(2\pi)$. In the $(\omega,k)$ subspace, the spectral function is typically sharply peaked around the dispersion relation \cite{Kamenev2011, Rammer2007, Berges2007, Gasenzer2009, Aarts2001}.

Applying Eq.~\eqref{eq:HamiltonPrinciple} to the resulting 1PI action yields the equation of motion for $\mathcal{A}_{k,t}$:
\begin{equation}
    (i\partial_t-\epsilon_k)\mathcal{A}_{k,t} = \int_{x}[\Sigma^R_{x,k,t}\mathcal{A}_{k,t}+\Sigma^K_{x,k,t}\tilde{\mathcal{A}}_{k,t}]-J_{k,t},
\end{equation}
where $\smash{\Sigma^{R/K}_{x,k,t} \equiv \int_\omega\Sigma^{R/K}_{x,k,t,\omega}B_{x,k,t,\omega}}$. Setting the sources and $\tilde{\mathcal{A}}_k$ to zero, and further using that $\mathcal{A}_k = \sqrt{2}\langle\hat{a}_k\rangle$, we finally recover Eq.~\eqref{eq:Akt}.

\subsection{Self-energy and kinetic equation}

The time evolution of $\langle\hat{a}_{k,t}\rangle$, Eq.~\eqref{eq:Akt}, is governed by the retarded self-energy $\Sigma^R_{x,k,t}$. By virtue of Eq. (\ref{eq:sigma}), the latter is solely determined by $\smash{\langle S_2^2e^{iS_{\text{int}}^\alpha}\rangle_0}$, which is quadratic in the fields $\mathcal{A}_k^i$. This expectation value corresponds to an infinite sum of diagrams that can, in principle, be evaluated perturbatively for a weakly interacting gas. The lowest-order term—known as the Born approximation—is obtained by expanding the exponential $e^{iS_{\text{int}}^\alpha}$ to zeroth order. However, for a gas at equilibrium, this approximation yields a vanishing self-energy \cite{Lange2012}. A nonzero result therefore requires a self-consistent resummation of the one-loop diagrams \cite{Andreev1980}. This resummation can be carried out either directly \cite{Keldysh1965, Duval2023, Buchhold2015} or via the Langreth rules \cite{Stefanucci2013, Kantorovich2020, Hyrkas2019}, leading to
\begin{align}
\label{eq:sigmaSCB}
&\Sigma^R_{x,k,t}\!=2i\int_{p,\nu,\omega}
\!\!\!\!B_{x,k,t,\omega}\Big[\Lambda_{p,k-p}^2G^K_{x,k-p,t,\omega-\nu}G^R_{x,p,t,\nu}\\ &+\Lambda_{k,p}^2(G^K_{x,k+p,t,\omega+\nu}G^A_{x,p,t,\nu}+G^K_{x,p,t,\nu}G^R_{x,k+p,t,\omega+\nu})\Big].\nonumber
\end{align}
This expression has the same structure as the Born-approximation expression \cite{Duval2023}, except that the Green’s functions entering it are the fully dressed ones rather than the free-space propagators. In the Wigner representation, they are given by $G^{R/A}_{x,k,t,\omega} = (\omega-\epsilon_{k}-\Sigma^{R/A}_{x,k,t})^{-1}$, where the self-energies are taken on-shell (i.e. at $\omega=c|k|$), and $G^K_{x,k,t,\omega} \simeq -iB_{x,k,t,\omega}(2n_{x,k,t}+1)$. Performing the integrals over $\nu$ and $\omega$, one finds:
\begin{align}
    &\Sigma^R_{x,k,t}\!=\! \int_p
    \Big[\frac{-4\Lambda_{k,p}^2(n_{x,p,t}-n_{x,k+p,t})}{\epsilon_{k+p}-\epsilon_{p}-\epsilon_{k}+\Sigma^R_{x,k+p,t}-\Sigma^A_{x,p,t}-\Sigma^A_{x,k,t}}\nonumber\\
        &-\frac{2\Lambda_{p,k-p}^{2^{\phantom{2}}}(n_{x,k-p,t}+n_{x,p,t}+1)}{\epsilon_{p}+\epsilon_{k-p}-\epsilon_{k}+\Sigma^R_{x,p,t}+\Sigma^R_{x,k-p,t}-\Sigma^A_{x,k,t}}\Big].
\end{align}
In this relation, the remaining integrals over $p$ are dominated by momentum regions corresponding to resonant scattering processes, i.e., those  
 that preserve both momentum and energy: $c|k|+c|p|=c|k+p|$ \cite{Andreev1980, Buchhold2015}.
Under this condition, the self-energy becomes purely imaginary, and one recovers
Eq. \eqref{eq:self-energyR}, if we use  that $\Sigma^A_{x,k,t} = \Sigma^{R*}_{x,k,t}=-\Sigma^{R}_{x,k,t}$.

We finally come to the kinetic equation \eqref{eq:kinetic} for the Wigner distribution $n_{x,k,t}$. This equation is obtained by combining the three Dyson equations for $G^R$, $G^A$ and $G^K$ encapsulated in the matrix relation \eqref{eq:Dyson}, with matrix elements expressed in Keldysh space as in Eq.~\eqref{eq:Gkktt}. By further using the anti-hermiticity of the Keldysh Green's function, one obtains a nonequilibrium fluctuation-dissipation relation for the Wigner distribution, see \cite{Kamenev2011, Rammer2007, Buchhold2015, Duval2023} for details: 
\begin{equation}
    (\partial_t+c\partial_x)n_{x,k,t} = \frac{i}{2}\Sigma^K_{x,k,t}-i(2n_{x,k,t}+1)\Sigma^R_{x,k,t}.
    \label{eq:noneqFDR}
\end{equation}
This equation involves the Keldysh self-energy $\Sigma^K$, which is computed from the self-consistent Born approximation, as for the retarded self-energy. The result is:
\begin{align}
   & \Sigma^K_{x,k,t} \!=\! -4\int_0^\infty
   \!\frac{dp}{\pi\rho_0}\frac{\Lambda_{k,p}^2(2n_{x,k+p,t}n_{x,p,t}\!+\!n_{x,k+p,t}\!+\!n_{x,p,t})}{\Sigma^R_{x,k+p,t}-\Sigma^A_{x,p,t}-\Sigma^A_{x,k,t}}\nonumber\\
    &-2\int_0^{k}\!\frac{dp}{\pi\rho_0}~\frac{\Lambda_{p,k-p}^2(2n_{x,k-p,t}n_{x,p,t}\!+\!n_{x,k-p,t}\!+\!n_{x,p,t}\!+\!1)}{\Sigma^R_{x,p,t}+\Sigma^R_{x,k-p,t}-\Sigma^A_{x,k,t}}.\nonumber
\end{align}
Inserting this relation and the corresponding retarded version \eqref{eq:self-energyR} into Eq. \eqref{eq:noneqFDR}, we finally obtain  Eq. \eqref{eq:kinetic}.

\section{Summary and conclusions}
\label{Sec:Conclusion}

In this work, we have provided a complete description of the relaxation dynamics of a weak periodic spatial modulation in a 1D Bose gas, from the quench protocol to the thermalization process. We showed that this relaxation is generally governed by a time-dependent phonon scattering rate $\gamma_{k,t}$, whose evolution is indirectly controlled by the dynamics of the phonon-mode populations generated by the quench. Only at long times, when these populations approach a thermal distribution, does $\gamma_{k,t}$ converge to its equilibrium value $\gamma_k$. We also found that this convergence is slow: the deviation $\gamma_{k,t}-\gamma_k$ decays algebraically in time, reflecting the algebraic slowing down of the phonon-distribution dynamics at long times due to energy conservation.

The periodic-potential quench considered here corresponds to a typical “spectroscopic’’ protocol in experiments, where the spatial period $2\pi/k_0$ of the imposed modulation allows direct access to the phonon relaxation rate at momentum $k_0$. Similar spectroscopic strategies have  been implemented in various platforms, such as cold atoms \cite{Stamper-Kurn1999, Ernst2010} or paraxial fluids of light \cite{Piekarski2021, Piekarski2025} to probe, for example,  the Bogoliubov dispersion or the structure factor in single- and two-component Bose fluids. Our work offers a theoretical guide for extracting the phonon scattering rate following a similar idea.

From a theoretical standpoint, the quantum hydrodynamics framework used in this work is rather general, as it can be extended to any (weak) initial spatial perturbation beyond the particular case of a periodic modulation considered here. The approach also extends straightforwardly to two-dimensional Bose gases, where the main modification is that a Born approximation is typically sufficient to compute the retarded self-energy, as long as interactions are weak. 
Importantly, the integrability of the microscopic Lieb–Liniger Hamiltonian \cite{Lieb1963a, Lieb1963b, Bulchandani2018, Bertini2016, Bouchoule2023} is effectively lost in our approach, as it is based on the effective coarsed-grained hydrodynamic Hamiltonian of Bose gases. This implies that our description, a priori, applies to situations where integrability is weakly broken at long times. This can happen for example in the presence of a shallow lattice (with lattice spacing well below the healing length) or by weak coupling to transverse degrees of freedom --- an effect inevitably present in 1D quantum-gas experiments \cite{Kinoshita2005, Langen2015, Dubois2024, Dubois2025}. Exploring the crossover from weak non-integrability to integrability in this setting is an interesting direction for future work.

\section*{Acknowledgements}

The authors are grateful to Adam Ran\c con, Gabriel Bonillo, H\'el\`ene Perrin, Isabelle Bouchoule, Maxime Jacquet and Quentin Glorieux for helpful discussions. NC acknowledges the financial support of Agence Nationale de la Recherche (ANR), France, under the Grants No. ANR-24-CE30- 6695 FUSIoN and Grant No ANR-23-
PETQ-0001 Dyn1D France 2030.

\appendix
\section{Post-quench phonon distribution}
\label{Sec:appendixA}

In this appendix, we derive the post-quench momentum distribution, Eq. \eqref{eq:pq_mom} of the main text. To this aim, we  compute the post-quench operator $\hat{a}_{k,\delta t}$ from the Heisenberg equation of motion 
$\smash{\partial_t\hat{a}_{k,t} = i[\hat{H},\hat{a}_{k,t}]}$ with $\hat{H} = \hat{H}_0+\hat{H}_{\text{int}}+\hat{H}_{\text{pot}}$, using a simple truncation scheme valid for a brief quench. As explained in the main text, if one drops the cubic part $\hat{H}_{\text{int}}$, the equation can exactly be solved, yielding Eq. \eqref{eq:apostquench} as a solution for $\hat{a}_{k,\delta t}$.
Now considering the cubic interactions, one has to compute the additional commutator:
\begin{equation}
    [\hat{H}_{\text{int}},\hat{a}_k] = 2\int_p\Lambda_{k,p}\hat{a}^\dagger_p\hat{a}_{p+k}^{\phantom{\dagger}}+\int_p\Lambda_{p,k-p}\hat{a}_p^{\phantom{\dagger}}\hat{a}_{k-p}^{\phantom{\dagger}}.
\end{equation}
At short time, this commutator can be approximated by writing $\hat{a}_k=\langle\hat{a}_k\rangle+\delta \hat{a}_k$ and linearizing with respect to $\delta \hat{a}_k$. This leads to
\begin{align}
    [\hat{H}_{\text{int}},\hat{a}_k] \simeq&~2\int_p\Lambda_{k,p}\langle\hat{a}^\dagger_p\rangle\langle\hat{a}_{p+k}\rangle+\int_p\Lambda_{p,k-p}\langle\hat{a}_p\rangle\langle\hat{a}_{k-p}\rangle \nonumber\\
    &+~2\int_p\Lambda_{k,p}\left(\langle\hat{a}^\dagger_p\rangle\delta \hat{a}_{p+k}+\langle\hat{a}_{p+k}\rangle\delta \hat{a}^\dagger_p\right)\nonumber \\
    &+\int_p\Lambda_{p,k-p}\left(\langle\hat{a}_p\rangle\delta \hat{a}_{k-p}+\langle\hat{a}_{k-p}\rangle\delta \hat{a}_p\right).
    \label{eq:CommutatorHintak}
\end{align}
The zeroth order terms in the first line give rise to sub-leading corrections to the field expectation value \eqref{eq:apostquench}, without any impact on the connected correlator $n_k$. We thus drop them in the following.
To evaluate the momentum integrals in Eq. (\ref{eq:CommutatorHintak}), we substitute each field expectation value by their lowest order expression \eqref{eq:meanapostquench}, and assume a coherent evolution of the same form as Eq. \eqref{eq:apostquench} for the time dependence of the off-diagonal operators $\delta\hat{a}_{p,t}$ and $\delta\hat{a}_{k\pm p,t}$. This leads to
\begin{equation}
\begin{split}
    [\hat{H}_{\text{int}},\hat{a}_k] \simeq&~\frac{V_0}{\epsilon_{k_0}}\sqrt{s_{k_0}}\big[(1-e^{-i\epsilon_{k_0}t})(\hat{a}^\dagger_-+\hat{a}^\dagger_+) \\
    &+2(1-\cos\left(\epsilon_{k_0}t\right))(\hat{a}_-+\hat{a}_+)\big],
\end{split}
\end{equation}
where we introduced $\smash{\hat{a}_\pm\equiv\Lambda_{k,\pm k_0}\delta\hat{a}_{k\pm k_0,0}e^{-i\epsilon_{k\pm k_0}t}}$ and $\smash{\hat{a}^\dagger_\pm\equiv\Lambda_{k,\pm k_0}\delta\hat{a}^\dagger_{-k\mp k_0,0}e^{i\epsilon_{k\pm k_0}t}}$.
Inserting this in the Heisenberg equation of motion and expanding the interacting correction for short quench durations, $\epsilon_k\delta t\ll1$, we obtain
\begin{align}
    \hat{a}_{k,\delta t} &=\langle\hat{a}_{k,\delta t}\rangle+\hat{a}_{k,0}e^{-i\epsilon_k \delta t} \\
    &+\frac{\varepsilon\sqrt{|k|}}{|k_0|}\left[\sqrt{|k-k_0|}\hat{a}^\dagger_{k_0-k,0}+\sqrt{|k+k_0|}\hat{a}^\dagger_{-k_0-k,0}\right]\nonumber
\end{align}
with $\varepsilon \equiv 3V_0(\epsilon_{k_0}\delta t)^2/(16g\rho_0)$. 
From this result, the computation of $n_{k,\delta t} \equiv \langle\hat{a}^\dagger_{k,\delta t}\hat{a}_{k,\delta t}\rangle_c$ is straightforward and yields, up to an irrelevant divergent contribution corresponding to a zero-point energy shift, Eq.~\eqref{eq:pq_mom} of the main text.


\begin{thebibliography}{99}
\bibitem{footnote}
In principle, the total energy also contains a mean-field term
$\int_kc|k|\langle\hat{a}^\dagger_k\rangle\langle\hat{a}_k\rangle$. However, for a weakly interacting Bose gas, this contribution is subleading—typically of order $\sim 1$ compared to the fluctuation term, $\int_kc|k| n_k\sim1/\sqrt{g}$.
\bibitem{footnote2}
Strictly speaking, at short times the expansions (\ref{eq:perturbationnsigma}) are not valid at large momentum. Nevertheless, the thermalization process governed by equations (\ref{eq:deltagamma}) and (\ref{eq:deltan}) primarily involves redistribution at low momenta, so that this inaccurate description has no significant impact.
\end{thebibliography}
\end{document}